\documentclass[runningheads]{llncs}

\usepackage{tabularx}
\usepackage{graphicx} %package to manage images
\usepackage{subcaption}
\graphicspath{ {./figures/} }

\begin{document}
\newcommand{\qa}[1]{{\color{blue}#1}}
\title{Orchestrating Collaborative Cybersecurity: A Secure Framework for Distributed Privacy-Preserving Threat Intelligence Sharing}%\\
%\small{Confidential - Article under review process - Please do not disseminate}}


% Juan Ramon Troncoso-Pastoriza juan@tuneinsight.com
% Alain Mermoud 
% Romain Bouyé romain@tuneinsight.com
% Francesco Marino francesco@tuneinsight.com
% Jean-Philippe Bossuat jean-philippe@tuneinsight.com
% Vincent Lenders 
% Jean-Pierre Hubaux jean-pierre.hubaux@epfl.ch (double affiliation: EPFL, Tune Insight SA)

%\author{Juan R. Troncoso-Pastoriza}
%\email{juan@tuneinsight.com}

\titlerunning{Orchestrating Collaborative Cybersecurity}
\author{Juan R. Trocoso-Pastoriza\inst{1} \and Alain Mermoud\inst{2} \and Romain Bouyé\inst{1} \and Francesco Marino\inst{1} \and Jean-Philippe Bossuat\inst{1} \and Vincent Lenders\inst{2} \and Jean-Pierre Hubaux\inst{1}\inst{3}}
\authorrunning{Troncoso-Pastoriza et al.}
\institute{Tune Insight SA, Switzerland \\
%\url{https://tuneinsight.com/}\\
\email{first@tuneinsight.com} \and Cyber-Defence Campus, armasuisse Science and Technology \\\email{first.last@armasuisse.ch} \and École polytechnique fédérale de Lausanne (EPFL)\\ \email{first.last@epfl.ch}}

%\author{Name}
%\affiliation{%
  %\institution{Cyber-Defence Campus, armasuisse Science and Technology}
 % \streetaddress{xxx}
 % \city{xxx}
 % \country{Switzerland}}
%\email{xxx}
%\renewcommand{\shortauthors}{xxx}

\maketitle

%%%While there is no formal page limit for DTRAP articles, we expect most submissions to be between 10 and 25 journal pages, with 30 pages being a soft upper limit.
\begin{abstract}
Cyber Threat Intelligence (CTI) sharing is an important activity to reduce information asymmetries between attackers and defenders. However, this activity presents challenges due to the tension between data sharing and confidentiality, that result in information retention often leading to a free-rider problem. Therefore, the information that is shared represents only the tip of the iceberg. Current literature assumes access to centralized databases containing all the information, but this is not always feasible, due to the aforementioned tension. This results in unbalanced or incomplete datasets, requiring the use of techniques to expand them; we show how these techniques lead to biased results and misleading performance expectations. We propose a novel framework for extracting CTI from distributed data on incidents, vulnerabilities and indicators of compromise, and demonstrate its use in several practical scenarios, in conjunction with the Malware Information Sharing Platforms (MISP). Policy implications for CTI sharing are presented and discussed. The proposed system relies on an efficient combination of privacy enhancing technologies and federated processing. This lets organizations stay in control of their CTI and minimize the risks of exposure or leakage, while enabling the benefits of sharing, more accurate and representative results, and more effective predictive and preventive defenses.
\end{abstract}


\keywords{cyber threat intelligence, information sharing, encrypted processing, multiparty homomorphic encryption, distributed cyber intelligence, MISP, cybersecurity}


\section{Introduction}
In the current interconnected world, the number of new threats and incident indicators keeps constantly increasing, to the point that it is impossible to adapt the detection and mitigation systems without an updated and comprehensive knowledge base that can be used to decipher the patterns of the incidents and train advanced models that can predict and detect them.

Today, there are many easy and cheap ways to share information: instant messaging, chat rooms, forums, emails, etc. However, despite the many technical solutions available, previous work has shown that cyber information sharing remains at sub-optimal level to deliver its full potential~\cite{mermoud_three_2019}. In fact, studies have shown that cyber information sharing is primarily a problem of (bad) human behavior and misaligned incentives~\cite{bohme_back_2016,bohme_economics_2013,falco2019cyber,david2020knowledge}. Existing research has shown that the human motivation to engage into cyber information sharing would be higher if the intelligence production is realized in a decentralized and privacy-preserving way~\cite{mermoud_incentives_2018} and under a public-private governance~\cite{mermoud_governance_2019} institutional design. This theoretical ideal is, however, difficult to implement in practice because of diverging interests and incentives between institutions ~\cite{anderson2006economics,anderson2001information}. 

Current efforts for sharing cyber threat intelligence, CTI, (e.g., the Malware Information Sharing Platform, MISP\footnote{\url{https://misp-project.org}}~\cite{wagner_misp_2016}), work on a centralized or replicated database, where all the participating organizations have to upload their threat data. As cyber information is often extremely sensitive and confidential, such efforts introduce a trade-off between the benefits of improved threat response capabilities and the drawbacks of disclosing national-security-related information to foreign agencies or institutions. This normally resolves in retention of the effectively shared information (aka the free-rider problem)~\cite{mermoud_share_2019}, which considerably limits the efficiency of collective action for tackling time-critical cybersecurity threats ~\cite{gillard2022efficient}. 

As stated in the white report produced by the World Economic Forum (WEF) on Cyber Information Sharing~\cite{WEF2020}, \textit{``information sharing is critical for empowering the global ecosystem to move from individual to collective cyber resilience''}. The WEF report also highlights the effects that the current global COVID-19 pandemic has had on pushing and speeding up the current digital transformation, hence exacerbating the cybersecurity challenges that existed before. Within this landscape, there is an urgent need for trusted, secure, and scalable cyber information sharing as an enabler of the global cybersecurity community, to (a) gather deeper insights on strategic, operational and technical information on cyber threats and risks (with examples in consortia such as FS-ISAC, the Cyber Threat Alliance -CTA-, CiviCERT, MM-ISAC, and the Telecommunication Information Sharing and Analysis Centre -T-ISAC), and to (b) drive collective investigations and actions to address cybercrime (with examples in initiatives such as the European Cybercrime Center -EC3-, the National Cyber-Forensics and Training Alliance -NCFTA-, Microsoft’s Digital Crime Unit -DCU-, or the Cyber Defense Alliance -CDA).

%Moreover, during the course of the aforementioned digital transformation process, new technologies such as machine learning and artificial intelligence are gaining an increasing impact and traction for enhancing the effectiveness of data sharing, and should be definitely considered as part of global cyberdefense strategies. This is part of the so-called Fourth Industrial Revolution, which involves highly relevant legal and technological challenges, among which trust and privacy play a prominent role.

The WEF report on Cyber Information Sharing also identifies the combination of Artificial Intelligence (AI)/Machine Learning (ML) and Privacy Enhancing Technologies (PETs) as the main enabler of the shift towards a new information sharing paradigm that can respond to the current cyberdefense challenges. AI and ML can address the scalability limitations of the current approaches to attack diagnosis by manual analysis of highly technical data, by automating these processes. This is achieved by introducing reliable and interpretable algorithms with high accuracy levels that can enable a more streamlined governance, decision-making and operating procedures. Nevertheless, in order to be effective, AI/ML has to be paired with appropriate PETs such as encrypted computation and differential privacy, that enable secure and federated data analysis, secure data linkage, secure search, and, in general, privacy-preserving machine learning with \textit{``encrypted AI models, protecting the model itself while preserving accuracy''}~\cite{WEF2020}.

All these considerations, already argued and justified from the privacy and security community, have been now articulated at the level of the WEF, with a huge impact and repercussion on cyberdefense stakeholders. A proof of this trend is the existence of current prototypes and test pilots, such as the CDA network of financial institutions in the United Kingdom, aimed at identifying and disrupting cybercrime activities in the financial sector by combining encryption and federated learning. These pilots prove that it is possible to preserve confidentiality and privacy while enabling timely responses to detect and deter malicious activities, and improving attribution and case building by banks and law enforcement. This is also a remarkable indicator of the timeliness of this work and the urgent need to expand and exploit the use of cryptographic techniques to enable efficient and secure insight sharing of CTI without transfer or disclosure of confidential-data.

%Supported by our know-how in securing statistical analyses in operational settings and highly-regulated environments (e.g., the MedCo~\cite{MEDCO} project in the domain of personalized health), and on a first investigation phase carried out in Q4 2020, in this work we propose a solution to enable a group of organizations to make use of their sensitive cyber-threat intelligence data by means of a framework and scalable software capable of orchestrating meaningful secure collaborations. We have identified mission-critical use cases that would inform the participating organizations about key-metrics on the gathered threat intelligence, and provide powerful analytical tools to improve on early detection and effective response to threats.

In this context, the target of this work is to resolve the cybersecurity information sharing trade-off by enabling more accurate insights on larger amounts of more relevant and confidential collective cyber threat intelligence. To the best of our knowledge, our work is the first to address this gap, by providing an applied solution for distributed privacy-preserving threat intelligence sharing.

\subsection*{Contributions}
\begin{figure}
\centering
    \includegraphics[width=0.7\textwidth]{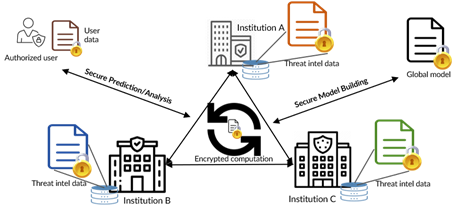}
    \caption{Cyber threat intelligence secure model building and prediction: data and models remain encrypted during computation.}
    \label{fig:threatintelmmodel}
\end{figure}

In order to fulfil the aforementioned target, this work presents a secure cyber threat intelligence sharing platform that offers provable technological guarantees that authorized users of the platform can only get access to the global insights (cyber threat models) built on the whole network data, whereas no access or transfer is granted on the local contributed data, which remains under the control of its source institution. For this, we present the following contributions, graphically depicted in Figure~\ref{fig:threatintelmmodel}:
\begin{itemize}
    \item We design and implement a distributed CTI sharing system without a centralized database (each institution keeps full control over their data records, which never leave their security perimeter).
    \item We integrate a combination of federated learning and cryptographic techniques (homomorphic encryption, multiparty computation, differential privacy) based on the paradigm of multiparty homomorphic encryption~\cite{Mouchet2020}, that make it possible to efficiently and scalably compute aggregate statistics and machine learning models on encrypted distributed data, and enable either a secure release of either the model, or just predictions produced by the model (model-as-a-service)~\cite{froelicher_scalable_nodate,POSEIDON}. Our implementation uses the open-source Lattigo library~\cite{lattigo}. %according to the system model depicted in Figure~\ref{fig:threatintelmmodel}.
    \item We exemplify and evaluate our system in three representative CTI sharing scenarios, where we compare it with prior work and show the challenges faced when the data from a single central database is not enough, and the advantages of our encrypted framework for overcoming those challenges and solve the CTI sharing trade-off.
\end{itemize}

\section{Privacy risks of federated CTI processing and countermeasures}
In this section, we briefly survey some of the recent applications of federated machine learning tools to cyber threat intelligence, highlighting the main privacy and leakage issues they face. Next, we provide a brief introduction to the different privacy-preserving approaches that can be used to address these issues, pointing out their pros and cons.

The use of machine learning and artificial intelligence for processing cyber threat intelligence has seen a considerable growth in recent years, in binary analysis \cite{Bao14,Shin15,Chua17,raff_malware_2017,vinayakumar_robust_2019}, 
%Tiffany Bao, Jonathan Burket, Maverick Woo, Rafael Turner, David Brumley, “Byteweight: Learning to recognize functions in binary code.” In 23rd USENIX Security Symposium (USENIX Security 14) (San Diego, CA, Aug. 2014), USENIX Association, pp. 845–860. Available online: https://www.usenix.org/conference/usenixsecurity14/technical-sessions/presentation/bao
%E. C. R. Shin, D. Song, and R. Moazzezi, “Recognizing functions in binaries with neural networks,” in 24th USENIX Security Symposium (USENIX Security 15). Washington, D.C.: USENIX Association, 2015, pp. 611–626. Available online: https://www.usenix.org/conference/usenixsecurity15/technical-sessions/presentation/shin
%Z. L. Chua, S. Shen, P. Saxena, and Z. Liang, “Neural nets can learn function type signatures from binaries,” in 26th USENIX Security Symposium (USENIX Security 17). Vancouver, BC: USENIX Association, 2017, pp. 99–116. Available online: https://www.usenix.org/conference/usenixsecurity17/technical-sessions/presentation/chua
%E. Raff, J. Barker, J. Sylvester, R. Brandon, B. Catanzaro, and C. Nicholas, “Malware Detection by Eating a Whole EXE,” ArXiv e-prints, Oct. 2017. (The Workshops of the Thirty-Second AAAI Conference on Artificial Intelligence)
%Vinayakumar, R., Mamoun Alazab, K. P. Soman, Prabaharan Poornachandran, and Sitalakshmi Venkatraman. "Robust intelligent malware detection using deep learning." IEEE Access 7 (2019): 46717-46738.
traffic analysis \cite{piskozub_malphase_2021} and network security~\cite{batchu_generalized_2021,fotiadou_incidents_2020}. Whereas most of the works in this area focus on centralized data, there have been some approaches towards federated processing. As a survey example, a recent work on using ML for intrusion detection at the IoT edge~\cite{wang_iot_edge_2019} presents an interesting feasibility analysis of ML at the edge, and a survey on ML algorithms used for security problems (e.g., intrusion detection) at the edge. The authors compare the suitability of supervised and unsupervised ML algorithms (decision trees/random forests, support vector machines, K-means, K-dimensional tree, DBSCAN, deep neural networks) to different cybersecurity issues (traffic data analysis, real-time response, time-series data analysis). They perform a qualitative comparison in terms of computation, memory footprint, accuracy and storage requirements of the analyzed algorithms. A relevant conclusion is that neural networks are comparatively adequate for time-series data analysis, and they can be implemented in a distributed (federated) fashion. %The results corroborate the suitability of the tools, proposed by our lab (EPFL/LDS) and industrialized by Tune Insight, for their use in cyber threat scenarios.

%It must be noted that the aforementioned work does not focus on privacy or confidentiality, but we include it here because there is currently a trend of platforms and systems that, as mentioned by the authors, use federated learning and edge computing as a privacy-preserving technique, as it helps in reducing the dimensionality of the data that leaves one institution; 
Whereas federated learning in this context can eventually reduce the amount of transferred data and limit it to only aggregate information instead of individual records or raw data, it has been shown in recent works that this aggregate data can also be subject to successful inference and re-identification attacks, enabling the extraction of parties inputs~\cite{orekondy19,hitaj17,wang19,zhu19} 
%Orekondy, Tribhuvanesh, Bernt Schiele, and Mario Fritz. 2019. “Knockoff Nets: Stealing Functionality of Black-Box Models.” In The IEEE Conference on Computer Vision and Pattern Recognition (CVPR).
%B. Hitaj, G. Ateniese, and F. Perez-Cruz. Deep models under the GAN: Information leakage from collaborative deep learning. In Proceedings of the 2017 ACM SIGSAC Conference on Computer and Communications Security, CCS ’17, page 603–618, New York, NY, USA, 2017. Association for Computing Machinery.
%Z. Wang, M. Song, Z. Zhang, Y. Song, Q. Wang, and H. Qi. Beyond inferring class representatives: User-level privacy leakage from federated learning. In IEEE INFOCOM 2019 - IEEE Conference on Computer Communications, pages 2512–2520, 2019.
%L. Zhu, Z. Liu, and S. Han. Deep leakage from gradients. In H. Wallach, H. Larochelle, A. Beygelzimer, F. dÁlché-Buc, E. Fox, and R. Garnett, editors, Advances in Neural Information Processing Systems 32, pages 14774–14784. Curran Associates, Inc., 2019.
or performing membership inference~\cite{nasr19,melis19}, 
%Nasr, Milad, Reza Shokri, and Amir Houmansadr. 2019. “Comprehensive Privacy Analysis of Deep Learning: Passive and Active White-Box Inference Attacks against Centralized and Federated Learning.” 2019 IEEE Symposium on Security and Privacy (SP), 739–53.
%L. Melis, C. Song, E. De Cristofaro, and V. Shmatikov. Exploiting unintended feature leakage in collaborative learning. In 2019 IEEE Symposium on Security and Privacy (SP), pages 691–706, 2019
raising concerns also from a regulatory perspective (in terms of the European General Data Protection, GDPR) on the use of federated learning alone as a privacy mechanism~\cite{truong20}. 
%Nguyen Truong, Kai Sun, Siyao Wang, Florian Guitton, Yike Guo. “Privacy Preservation in Federated Learning: Insights from the GDPR Perspective”	ArXiV CoRR cs.CR arXiv:2011.05411. 2020.
This problem is worsened in settings where the number of features is comparable (or larger) to the number of data records (e.g., deep learning), and also in iterative approaches, where more (aggregate) data is transferred at each iteration. There, the ``dimensionality reduction'' as a privacy argument in favor of federated learning becomes unrealistic.

Privacy works unrelated to CTI traditionally address this problem by resorting to differential privacy techniques~\cite{milletari19,shokri15,mcmahan18}. 
%W. Li, F. Milletarì, D. Xu, N. Rieke, J. Hancox, W. Zhu, M. Baust, Y. Cheng, S. Ourselin, M. J. Cardoso, and A. Feng. Privacy-preserving federated brain tumour segmentation. In H.-I. Suk, M. Liu, P. Yan, and C. Lian, editors, International Workshop in Machine Learning in Medical Imaging (MLMI). Springer, 2019.
%R. Shokri and V. Shmatikov. Privacy-preserving deep learning. In ACM Conference on Computer and Communications Security (CCS), 2015.
%H. B. McMahan, D. Ramage, K. Talwar, and L. Zhang. Learning differentially private recurrent language models. In International Conference on Learning Representations, 2018
But training robust and accurate models requires high privacy budgets, and as such, utility is impacted~\cite{shmatikov19}, 
%Eugene Bagdasaryan, Andreas Veit, Yiqing Hua, Deborah Estrin, Vitaly Shmatikov. 2019. "How To Backdoor Federated Learning", CoRR arXiv:1807.00459
and the level of privacy achieved in practice remains, in general, unclear~\cite{jayaraman19}. 
%B. Jayaraman and D. Evans. Evaluating differentially private machine learning in practice. In USENIX Security, 2019.
The available cryptographic approaches to address this leakage are briefly presented in the following paragraphs.

%Other approaches comprise the use of cryptographic mechanisms in combination with federated learning. This combination avoids the accuracy penalty incurred by differential privacy, while keeping the performance and distribution properties of federated learning, and guaranteeing a provable level of confidentiality.There are several

Privacy solutions based on \emph{trusted execution environments} (TEEs) rely on a trusted tamper-proof hardware element deployed within an untrustworthy environment; they define protocols to deliver secret keys to the enclave so that it is possible to send to it encrypted data to be processed; a private region of memory contains encrypted data and code (enclave) and it is decrypted on the fly and processed by the CPU. Key delivery is usually paired with an attestation process that validates the code running in the enclave. The security is based on the guarantees of the key delivery and attestation protocols, and on the assumption that the hardware realization (and firmware implementation) of the tamper-proof module cannot be breached.
An example of the use of TEEs in cybersecurity is TrustAV~\cite{trustav}. 
%Dimitris Deyannis, Eva Papadogiannaki, Giorgos Kalivianakis, Giorgos Vasiliadis, and Sotiris Ioannidis. 2020. “TrustAV: Practical and Privacy Preserving Malware Analysis in the Cloud.” In Proceedings of the Tenth ACM Conference on Data and Application Security and Privacy (CODASPY '20). Association for Computing Machinery, New York, NY, USA, 39–48. DOI:https://doi.org/10.1145/3374664.3375748
TrustAV is a cloud-based malware detection solution that offloads the processing of malware analysis to a remote server, where it is executed entirely inside hardware supported secure enclaves. The tool also needs memory optimization techniques to reduce the required enclave memory, a limiting factor for malware analysis executed in secure enclave environments.

It must be noted that TEEs are centralized (all data is sent to a central enclave), and it may not be practical or feasible for cross-jurisdictional data sharing. Furthermore, enclaves are hardware- and infrastructure-dependent, and they require trust in the platform manufacturer. Moreover, the recent work of hardware security researchers is producing a constant stream of new vulnerabilities in the current most widespread and practical implementations of secure enclaves (in particular, Intel SGX). Vulnerability patching and updates in TEE-based systems is much less flexible and costly than in software-based systems. Due to these considerations, we will focus only on software-based approaches. %For a full comparison between software and hardware privacy-techniques, we refer the reader to\cite{HdrwVsSftwrTI2021}.

\emph{Homomorphic encryption (HE)} is a special form of encryption that enables calculations on encrypted data without decrypting it first. Therefore, it preserves the confidentiality of input data with respect to an untrustworthy computation environment, at the cost of an increased computational complexity. In the cybersecurity field, an example is SCRAM~\cite{Castro2020SCRAM} (Secure Cyber Risk Aggregation and Measurement), that enables multiple entities to compute aggregate cyber-risk measures without disclosing the individual sensitive data from each of the institutions.  %The work presents results on “(1) benchmarks of the adoption rates of 171 critical security measures and (2) links between monetary losses from 49 security incidents and the specific sub-control failures implicated in each incident.”
The system is a relatively simple prototype that requires manual intervention in air-gapped terminals to input the data, and the performed computations are restricted to counts and additions, with a very ad-hoc setting (data-dependent predefined high-level statistics). It must be noted also that this work uses a central entity for key and data aggregation and the analysis suffered from data quality problems (one of the firms had wrongly formatted the input data). Nevertheless, it shows the feasibility of a homomorphic encryption-based solution to enable the computation of simple but relevant statistics on distributed cyber threat data.

\emph{Secure Multiparty Computation (SMC)} solutions are based on secure interactive protocols that enable several parties to jointly compute a function on their respective inputs without revealing anything about their inputs besides what can be inferred from the output of the function. Unlike homomorphic encryption, they are computationally light, but they are usually bottlenecked by the used bandwidth and communication, and are thus usually limited to three or four parties.

\subsection*{Our approach}
The hybrid approach taken in this work is based on a family of techniques that we denote Multiparty Homomorphic Encryption (MHE)~\cite{Mouchet2020,lattigo}. This paradigm combines the strengths of homomorphic encryption (efficient communication, secure outsourced processing, hardware-agnosticism) and of secure multiparty computation (efficient computation, versatile functionality, hardware-agnosticism). The combination of these two techniques, when applied to a machine learning scenario where data is partitioned across several nodes leveraging the scalability of federated learning approaches result in an optimal solution that enables high efficiency, scalability, performance and accuracy, with low bandwidth usage and the same cryptographic security guarantees as homomorphic encryption alone provides, while avoiding its limitations and thwarting inference and reconstruction attacks that would be effective for federated learning approaches.

Furthermore, this combination of HE and SMC has been analyzed and proposed as applicable measures conducive to compliance with data protection regulatory frameworks such as the European GDPR, both as means to achieve security- and privacy-by-design and as supplementary measures for enabling international data transfers~\cite{scheibner21,corrales21}. 
%James Scheibner, Jean Louis Raisaro, Juan Ramón Troncoso-Pastoriza, Marcello Ienca, Jacques Fellay, Effy Vayena, Jean-Pierre Hubaux. "Revolutionizing Medical Data Sharing Using Advanced Privacy-Enhancing Technologies: Technical, Legal, and Ethical Synthesis" J Med Internet Res 2021;23(2):e25120 doi: 10.2196/25120
% Marcelo Corrales Compagnucci, Mateo Aboy, Timo Minssen. "Cross-Border Transfers of Personal Data After Schrems II: Supplementary Measures and new Standard Contractual Clauses (SCCs)". Nordic Journal of European Law Issue, Vol 4(2) 2021, https://doi.org/10.36969/njel.v4i2.23780
This is of utmost importance to certify the regulatory compliance of systems implementing MHE solutions, and to streamline the validation of our distributed secure analytics platform for cross-border collaborations.

\section{Analyzed use-cases of CTI sharing}
We analyze three different representative use cases of cyber intelligence sharing of increasing complexity, that exemplify the information retention problem. We show the relevance and beneficial impact of using our system on these scenarios. The first case involves the computation of aggregate statistics, whereas the second and third cases involve the training of machine learning models used for the detection and classification of cyber incidents and events:
\begin{enumerate}
    \item Global statistics of MISP events
    \item Training and prediction of threat level from MISP events
    \item Training and prediction of DDoS attacks
\end{enumerate}

For each of these use cases we take one representative prior work, that we use in the following sections as a baseline to highlight the challenges in previous attempts of addressing these scenarios, and to evaluate and compare the performance and the benefits of the proposed solution.

\subsection{Global statistics of MISP events}
In this scenario, a network of organizations record in MISP the detected intrusions and the type of malware detected by their antivirus or by using identification tools such as VirusTotal.

Each organization saves in their local MISP instance the detected events with the associated name of the identified malware, taken from a common taxonomy (exemplified in Table~\ref{tab:malwareaggreg}). These organizations do not want to share the recorded events, because they do not want to unveil the effect and impact of the intrusions they suffer.

\begin{table}
\caption{Example malware taxonomy used in the global statistics scenario.}
\label{tab:malwareaggreg}
\centering
\begin{tabular}{|l|l|}\hline
    \textbf{Event Date}&	\textbf{Malware name}\\\hline
    2020-08-10&	Locky\\\hline
    2020-08-12&	Emotet\\\hline
    2020-08-12&	Ryuk\\\hline
    2020-08-15&	WannaCry\\\hline
    2020-08-17&	Fireball \\\hline
    …&	…\\\hline
\end{tabular}
\end{table}

The different organizations would like to calculate statistics on the collective dataset, such as the global number of intrusion events per type of malware. This allows them to extract global statistics that can be helpful as input information for their SOC/SIEM systems.
By comparing the global number of intrusions with its own numbers, an organization can contextualize its position in the network, understand the scope of the attackers’ targets and the trends of the most prevalent threats, and adapt the security processes accordingly. For example, if ransomware appears as an increasing global threat, data backup and replication policies can be strengthened.

This use case is a generalization of the example proposed in~\cite{Castro2020SCRAM}. The use of MISP in our use case addresses the challenges of data homogenization and of automation of data collection, by providing a standardized common data source to the collective statistics system.

\subsection{Training and automatic prediction of threat level with MISP events}
Several organizations store information about indicators of compromise on their respective MISP databases. We have taken as an example the features stored in the CIRCL MISP\footnote{\url{https://www.circl.lu/services/misp-malware-information-sharing-platform/}}, shown in Table~\ref{tab:allcirclemispfeatures} in the Appendix.

%\begin{table}
%    \begin{tabular}{|l|l|l|}\hline
%        \textbf{Name}&\textbf{Description}\\\hline
%MISSING->Add table with the list of all used features (both section 6.2 and 6.3), and move to appendix. Romain
%    \end{tabular}
%    \caption{Taxonomy of the MISP data used for the threat level prediction scenario.}
%    \label{tab:mispfeatures}
%\end{table}

These fields are typically the result of a primary analysis of the event, but for this specific dataset they do not entail critically confidential information (such as local IPs or email addresses). In fact, the nature of the data shared in the CIRCL MISP is not the raw data from an attack but instead the report information after an initial analysis by an operator. Moreover, this data is put together with the purpose to be shared, and it is usually not deemed confidential. Nevertheless, this is a good example on how MISP databases could be enriched with more relevant, up-to-date and critical information that can be used to train a model that can better predict the threat level of a new detected indicator, without having to share the original information.

This scenario is a generalization of Fotiadou et al.'s work~\cite{fotiadou_incidents_2020}. The authors trained a machine learning model using data extracted from MISP; they use and compare several machine methods such as Multi-layer Perceptrons and Convolutional Neural Networks (CNN). Those models are used to predict the threat level of a network request.

\subsection{Training and detection of DDoS attacks}
In this scenario, the target is to train a model for detection of DDoS attacks. Multiple organizations collect data related to the historic data on DDoS attacks and of normal traffic in their networks, and they want to use this information to train a more accurate and generalizable collective model that can improve the detection of DDoS attacks. Obviously, the network traces from each organization represent confidential information that they are not willing to share or centralize.

This is a case that more naturally reflects the need for multi-party homomorphic encrypted analytics when the analyzed information cannot be shared between the participating organizations. During a Distributed-Denial-of-Service (DDoS) attack, a firewall system must distinguish benign from malignant traffic. This can be done by identifying patterns from private system information such as the IP addresses of the source and destination, the used port and protocol, and information about the TCP packet.

For this scenario, we take as a baseline a prior work by Batchu and Seetha~\cite{batchu_generalized_2021} for detection of DDoS attacks with a centralized dataset. %Batchu and Seetha~\cite{batchu_generalized_2021} demonstrate the feasibility of reliably identifying malign traffic using the public dataset CICDDoS2019 \footnote{Add ref}. This dataset contains this kind of confidential traffic information, but it has been made public, so the results are fully reproducible and comparable to a distributed scenario.
Batchu and Seetha~\cite{batchu_generalized_2021} proposed a generalized machine learning model for DDoS attacks detection. They use a novel automatic detection method to reduce the feature space and then they perform hyperparameter tuning to select the most appropriate parameters for different learning approaches. Then, they feed both the optimal features and hyperparameters to various supervised learning approaches (logistic regression, decision tree, gradient boost, k-nearest neighbor, and support vector machine). They demonstrate the feasibility of reliably identifying malign traffic using the public dataset CICDDoS2019 \footnote{\url{https://www.unb.ca/cic/datasets/ddos-2019.html}}. This dataset contains this kind of confidential traffic information, but it has been made public, so the results are fully reproducible and we can use them to compare them to a distributed scenario.

\section{System and threat model}
We briefly introduce here the stakeholders that partake in the cyber threat intelligence sharing scenario, together with the corresponding threat model that can be assumed for designing our secure architecture, and the system model.

\subsection{Stakeholders and threat model}
The stakeholders can be classified in three different categories or logical roles. The same stakeholder can play several overlapping roles within the infrastructure of a given organization.

\begin{itemize}
\item \emph{Data providers or data sources}: These are intelligence agencies and cyberdefense groups within networked organizations that are part of a MISP community and store their own collected cyber threat intelligence in a MISP instance. Operators or automated systems in those entities can act as data sources. The data sources are expected to be honest by providing real and trustworthy data (this assumption could be released in more strict scenarios).
\item\emph{Data processors or computing nodes}: A given institution can run their analysis workflows and computational processing of internally or externally generated data either in-house or relying on a private or a public cloud. The infrastructure running the analysis is considered semi-honest (honest but curious, or passively adversarial), as it provides the computing capabilities and should follow the protocols, but can try to infer additional information about the data and computation results to which it has access.
\emph{Data consumers}: In our scenarios, the role of data consumers comprises the same entities that can play the role of data providers. Within these organizations, data analysts or information security experts can query the cyber threat intelligence and request analyses and/or access to aggregated data. Automated tools can also be used to generate alerts or update the systems according to the detected threats. We assume that the data consumers are also semi-honest.
\end{itemize}

For the sake of simplicity, and following a win-win collaboration between $N$ networked cyberdefense groups and intelligence institutions, we consider that each organization acts both as data provider (of its internally generated intelligence) and consumer (of both internally and externally generated intelligence from the network), and it has (in-house or outsourced) computation capabilities to perform analyses on their data and to contribute to distributed analyses on collective network data. Therefore, we consider $N$ nodes (parties), under a passive-adversary (semi-honest) model with collusions of up to $N-1$ parties: i.e., all nodes follow the protocol, but up to $N-1$ nodes might share among them their inputs, and their observed intermediate and final results, to extract information about the other parties’ inputs through membership inference or federated learning attacks that our system has to prevent.

\subsection{System Model}
\begin{figure}
\centering
    \includegraphics[width=0.7\textwidth]{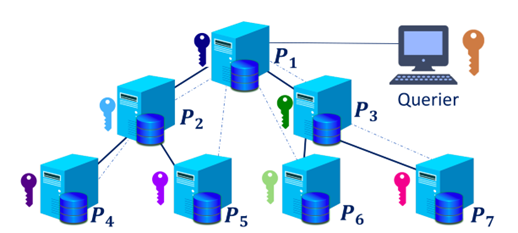}
    \caption{System Model comprising $N$ interconnected parties and a querier, each with a public-private key-pair. Each party plays the role of data provider and computing node.}
    \label{fig:systemmodel}
\end{figure}

We consider the setting depicted in Figure~\ref{fig:systemmodel}, with $N$ parties, each one locally holding its own labeled data, which can be represented by a flat $n\times m$ matrix $X_i$ ($n$ records and $m$ variables) and a label vector $y_i$ of length $n$. Each party owns a public-private key pair, and all of them use a collective public key (generated by aggregating all individual public keys) for encrypting the data used during the secure computation~\cite{Mouchet2020}. These parties enable the computation of aggregate statistics and the collective training and evaluation of machine learning models. The computation of aggregate statistics or a model training can be triggered by an authorized system user (querier). At the end of a training process, a querier - which can be one of the $N$ parties or an external entity - can request the evaluation of the trained model to obtain classification or prediction results $y_q$ on its input evaluation data $X_q$.
The parties involved in the training process are interested in preserving the confidentiality of their local data, the intermediate model updates, and the resulting models. The querier obtains aggregate statistics and classification/prediction results on a trained model and also requires that its evaluation data is kept confidential. We assume that the parties are interconnected and can be organized in a tree-network structure for more efficient communication, as shown in Figure~\ref{fig:systemmodel} (thick lines). However, our system is fully distributed and does not assume any hierarchy, therefore remaining agnostic of the network topology, e.g., we can consider a fully-connected network, or a star topology in which each party communicates with a central server (dotted lines in Figure~\ref{fig:systemmodel}).

\section{Proposed system architecture}
\label{sec:architecture}
\begin{figure}
\centering
    \includegraphics[width=1\textwidth]{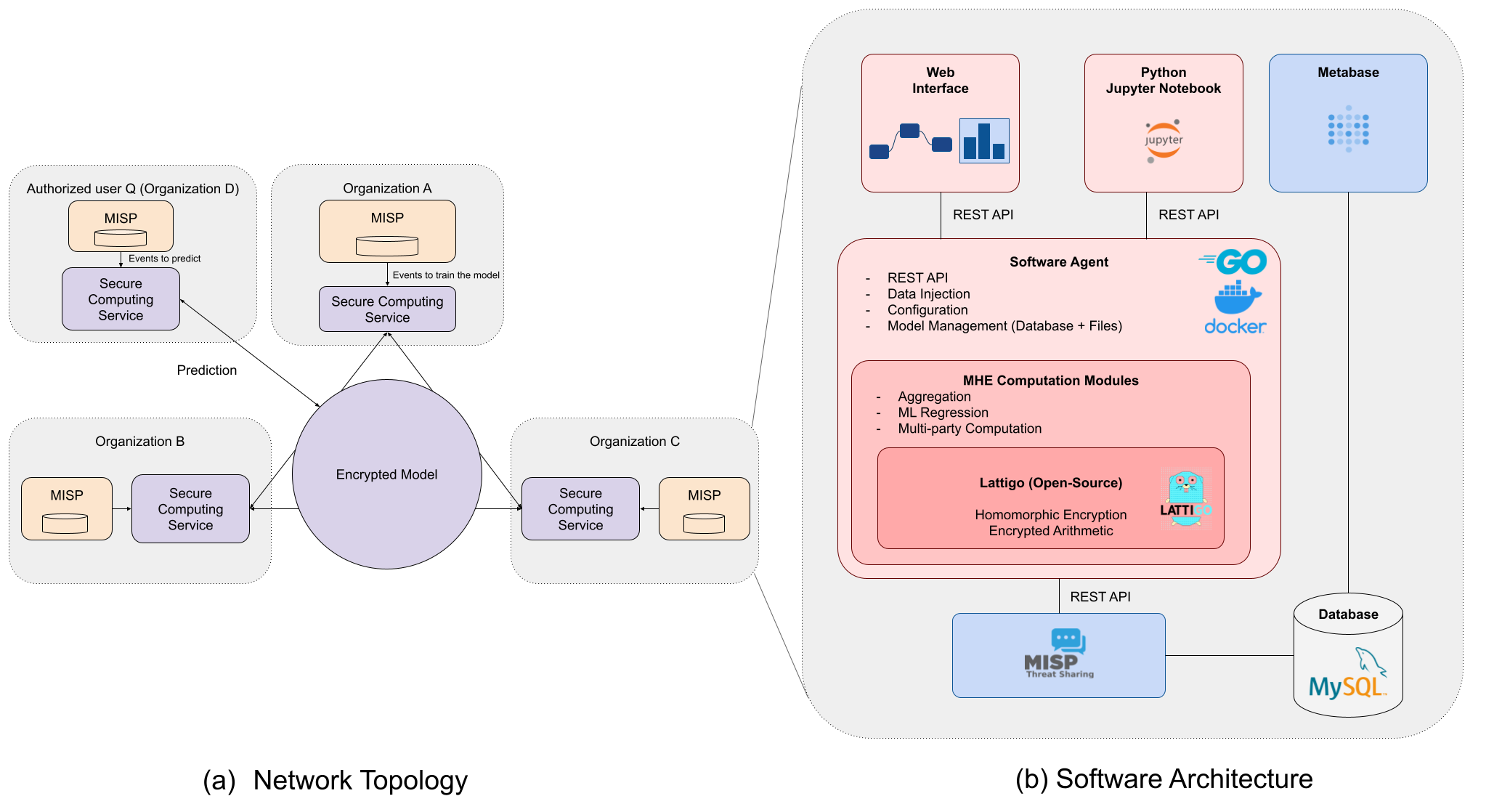}
    \caption{Conceptual diagram (a) of the secure computing platform, including the connected MISP instances at each organization and (b) the software components of a deployed node at each organization.}
    \label{fig:MISPmodel}
\end{figure}

Figure~\ref{fig:MISPmodel}(a) shows an example of the conceptual diagram of our secure computing service, with the integration of MISP backends at each organization. This figure shows three organizations (A, B and C) that set up a network for secure distributed training and evaluation of machine learning models based on MISP data. It is assumed they all have an instance of MISP running a priori, and that they have matching and interoperable data taxonomy and semantics. Authorized users can be affiliated with one of the collaborating organizations or with a third party (Organization D in the figure).

Each organization deploys an instance of our secure computing service that enables secure distributed cryptographic computations. This service ensures that no clear-text data is sent out of the organization, and, together with analogous instances running at the other organizations, it enables the computation of aggregate statistics, the training of machine learning models, and the evaluation of prediction/classification tasks using those models.

The software components shown in Figure~\ref{fig:MISPmodel}(b) are bundled in Docker images and organized in a single package. The package contains an installation script to easily load the images and configure the node. After the installation, each organization can run all component with Docker Compose. Metabase is used as a visualization tool for the data from MISP, and it is an optional component in the package. The software agent is responsible to securely pull, process and encrypt information from MISP without exposing it to non-authorized users.
The authorization and access right management is highly customizable and handled by an external authorization provider, described in section \ref{sec:keycloak}.
% Done. \qa{Romain: add more details about the packaging and deployment (Docker, automation of the network configuration, permissioning, key management - permissioning and key management are already added in the extension section so there can be a forward reference here,...)?}

Our secure software agent is connected to MISP and pulls events via its REST API. This approach is efficient for a small number of events (<10,000), and it is appropriate to regularly extract a relatively small amount of data from MISP, but it is not the most efficient to pull a large dataset, such as when exporting all the available events. For our study evaluation, we considered both the periodic data extraction and the dump of events data from each local MISP instance to a local instance of the data analytics software Metabase~\footnote{https://www.metabase.com/}, directly connected to the MISP MySQL database. This enables organizations to explore the data both using the user interface and directly running SQL queries.

\begin{figure}
\centering
    \includegraphics[width=0.7\textwidth]{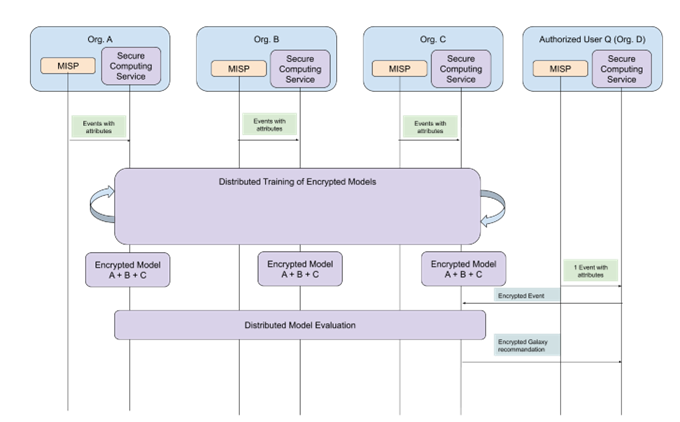}
    \caption{Sequence diagram of a typical secure computing process involving model training and recommendation based on MISP data.}
    \label{fig:MISPsequence}
\end{figure}

Figure~\ref{fig:MISPsequence} shows the same system model previously described with the added sequence of events, to detail how the system operates. We now detail the main functionalities: Model training, model evaluation, and aggregate statistics computation.

\begin{itemize}
    \item \emph{Model training}. An authorized user Q (potentially in a different organization ``D'') can query the network to train secure encrypted models on selected data. In order to ensure confidentiality, this data is encrypted locally within the organization before it is used by the secure computing service, and the resulting model will also be encrypted to avoid any leakage.
    \begin{enumerate}
        \item The authorized user selects what data from the network has to be used for training the model.
        \item When the training is triggered, the client notifies the local secure computing service, which starts a coordinated protocol with all the secure computing services in the network.
        \item The secure computing service at each node fetches the selected data from its corresponding local MISP instance (``events with attributes'').
        \item The distributed training is executed across the network, by interchanging encrypted aggregated data (under a collective public key) in a secure protocol executed and coordinated by the secure computing services. As a result of this process, all organizations obtain and store the encrypted model.
    \end{enumerate}

    \item \emph{Model evaluation}. Once a model is trained, the user can use a data set in her possession to request predictions/classifications or automated labeling with the encrypted model. The process does not reveal the model itself or the data that was used for training. Moreover, the confidentiality of the predictions/classifications is guaranteed by the fact that the result that is sent back to the user is encrypted with the user's key, without having ever been decrypted during computation.
    \begin{enumerate}
        \item An authorized user selects MISP data from her local MISP instance or from another local source to be used for prediction, classification or automated labeling.
        \item The user interface transparently encrypts this data (with a collective public key) and sends it to the local secure computing service in their organization.
        \item This encrypted data is input to a distributed prediction protocol across the network to compute the encrypted predictions from the model.
        \item The resulting encrypted predictions/labels then pass through an additional distributed protocol to be re-encrypted under the key of the user, making her or him the only person able to decrypt them.
        \item They are sent back to the user, who can now decrypt (transparently in the client application) and use the results.
    \end{enumerate}
    
    \item \emph{Aggregate statistics computation} Unlike the model training/evaluation, the computation of aggregate statistics is run in a single phase, which can be seen as the concatenation of model building and model evaluation, in which only the aggregate statistics are output as a result.
    \begin{enumerate}
        \item An authorized user selects a filtered subset of MISP data from the network to be used for computing aggregate statistics.
        \item The client notifies the local secure computing service, which starts a coordinated protocol with all the secure computing services in the network.
        \item The secure computing service at each node fetches the selected data from its corresponding local MISP instance (``events with attributes'').
        \item The distributed aggregation is executed across the network, by interchanging encrypted aggregated data (under a collective public key) in a secure protocol executed and coordinated by the secure computing services. As a result of this process, the aggregate statistics are obtained encrypted under the collective public key.
        \item The resulting aggregate statistics then pass through an additional distributed protocol to be re-encrypted under the key of the user, making her or him the only person able to decrypt them.
        \item They are sent back to the user, who can now decrypt (transparently in the client application) and visualize the statistics and/or enrich its local MISP instance with the newly gathered insight.
    \end{enumerate} 
    
\end{itemize}

\subsection*{Enabled machine learning models under MHE}
The used homomorphic cryptosystems in MHE enable the efficient computation of polynomial functions, which can be used to accurately approximate other functions. In order to deal with complex workflows, MHE leverages the combination of homomorphic encryption and secure multiparty computation. This enables an efficient and scalable computation of some machine learning algorithms in federated settings, including regression models~\cite{froelicher_scalable_nodate}, feed-forward neural networks~\cite{POSEIDON}, and time-to-event analysis~\cite{froelicher_naturecomms} among others.

In the rest of the work, we focus on logistic regression algorithms as the exemplified machine learning tool. In spite of its simplicity, it provides a valid comparison baseline and very satisfactory results for the addressed problems; additionally, it enables better explainability than other more complex models such as neural networks, and its computation is very efficient. Also, the obtained results can be extrapolated to more complex models.

It must be noted that, in order compute under encryption the activation functions used in the logistic regression training (Sigmoid functions), we approximate them by polynomials. The degree and coefficients of the polynomial can be optimized depending on the dataset as part of the parametrization for the encrypted logistic regression. This approximation introduces a controlled bounded modification in the results that does not affect convergence but that can have a small impact on the resulting precision, only noticeable when the dataset contains a significant amount of outliers. As part of the future work for the developed platform, it is foreseen to include an automatic parametrization and selection of the best approximation functions as part of the local data analysis before the data is ingested in the distributed training protocol.

\section{Evaluation Results}
In this section, we provide the evaluation results for the analyzed three cases. This evaluation is focused on the functionality, representativity, performance and accuracy achieved in these scenarios, each of them highlighting specific dimensions of these magnitudes. The first two cases exemplify simple functionalities that can serve to effectively enrich the collective intelligence in an automated and efficient way when compared with prior work in the literature aiming at similar scenarios, whereas the third use case is more complex and has more realistic and representative publicly available data, so we use it to fully analyze the problems of bias introduced by synthetic data and to establish a comparison in terms of accuracy and performance with respect to an ideal centralized solution and a distributed non-secure solution.

We have implemented the secure computing service shown in Section~\ref{sec:architecture} using Go, on top of the Lattigo library~\cite{lattigo}. The evaluation is run on instances of machines with 2x12 Intel Xeon cores@2.5 GHz and 256 GB of RAM. The developed solution is packaged in Docker containers and deployed in this infrastructure. Tests with a small number of nodes are performed on one node per machine, whereas large scale tests are performed by emulating nodes in virtual machines.

\subsection{Global statistics of MISP events}
Using Multiparty Homomorphic Encryption, it is possible to compute aggregate statistics from the collective datasets without having to share the original data. The MISP instances at each institution are therefore only used to structure and store the local data, but they do not replicate or transfer the data collected by each of them to the others. Our platform receives queries and aggregates the results under encryption to give back only the requested statistic without transferring any individual record.

\begin{figure}
\centering
    \includegraphics[width=\textwidth]{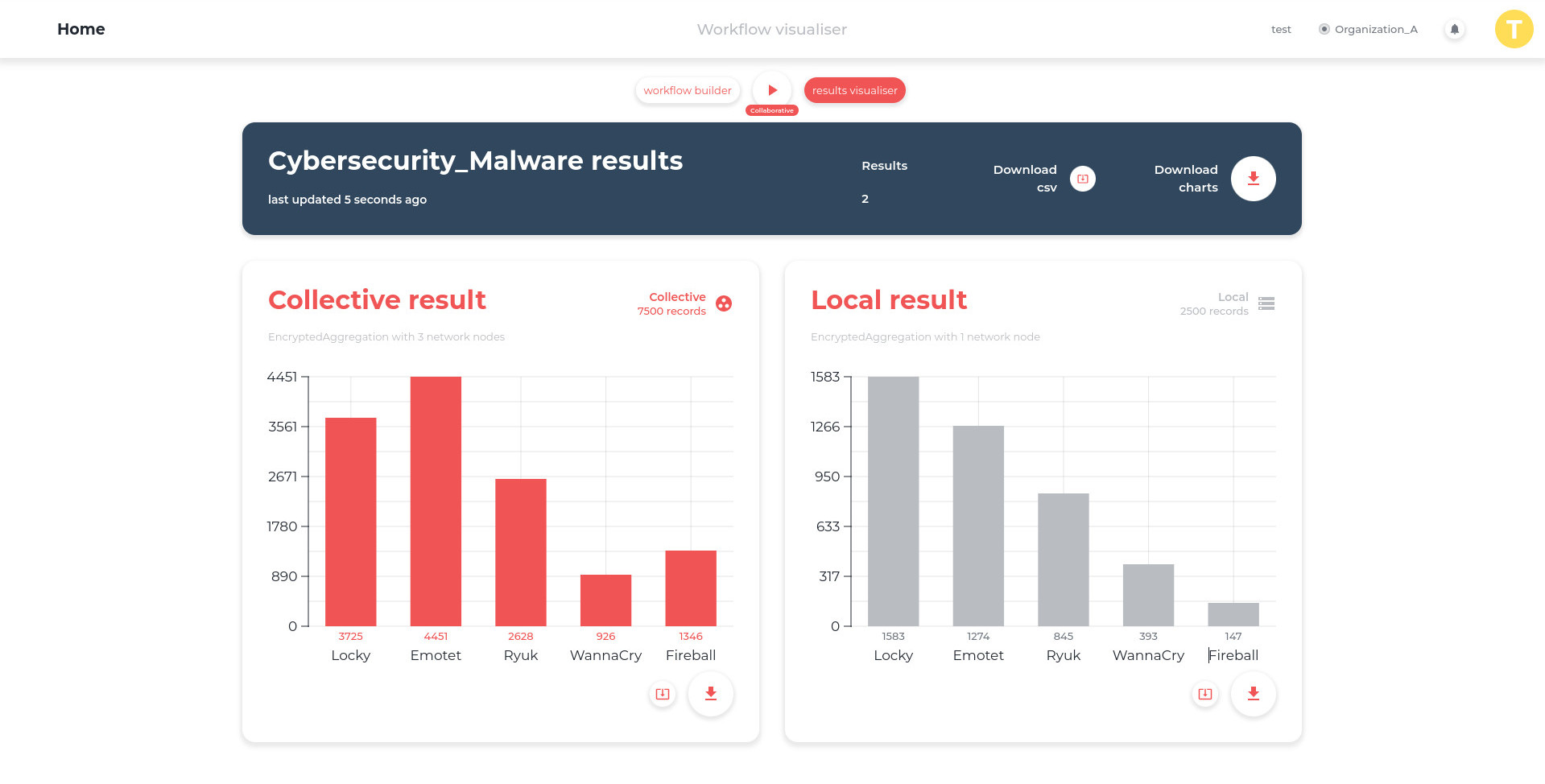}
    \caption{Example of aggregated results across the network presented in the system UI.}
    \label{fig:malwareaggreg}
\end{figure}

Figure~\ref{fig:malwareaggreg} shows a simple example with the histogram of the total number of malware events per type during a time period, visualized through our user interface (anonymized for the purposes of the review process). The used data for this example is synthetically generated. Regarding the measured performance, as most of the computation happens locally at the level of the MISP instances, the encrypted computation overhead is negligible with respect to the database response time.

This case is comparable to the example proposed in~\cite{Castro2020SCRAM}. As highlighted above, our system and its user interface shown in Figure~\ref{fig:malwareaggreg} enable a more standardized, streamlined and automated data collection, and an easy and direct way of defining one-shot or periodic retrieval of collective aggregate statistics from the locally populated MISP instances. Thanks to the versatility of MISP, this solution is also agnostic of the sector, and it is not only applicable to financial services.

\subsection{Training and prediction of threat level from MISP events}
In order to prepare and select the used features we employed two different methodologies.

Firstly, we ran an instance of MISP and synchronized it to pull events and attributes from the CIRCL MISP. We then exported 1,048,576 attributes related to network activity (containing URLs or IP addresses). Each attribute relates to an event with an assigned threat level ranging from 1 to 4. Using a selection of 39 features (see Appendix 1), we converted categorical data into binary columns and standardized it. We used 90\% of the dataset as input to train a logistic regression model using a stochastic gradient descent (SGD) algorithm, and the remaining 10\% (104,858 attributes) to test the model.

Secondly, we followed the same approach as in~\cite{fotiadou_incidents_2020}, by exporting from the local MISP MySQL databases a dataset of 100’490 events in total. For the data preparation process, we used a standard Python method with the Pandas and Scikit-learn libraries, run at the local execution environment. We converted the data into numerical entries, removed empty values and standardized it.

We tested the usefulness and effectiveness of Machine Learning algorithms such as linear and logistic regressions over data extracted from MISP, as tools to automate classification tasks. The used dataset consists of network traffic attributes related to events with an associated threat level. By selecting a set of features from those attributes, it is possible to effectively predict the threat level of the event.

\begin{figure}
    \centering
    \begin{subfigure}[b]{0.9\textwidth}
        \includegraphics[width=0.45\textwidth]{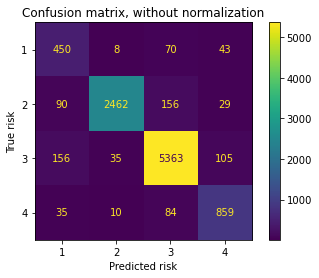} 
        ~
        \includegraphics[width=0.45\textwidth]{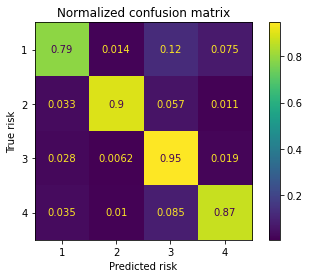}
        \caption{Results obtained using the methodology in~\cite{fotiadou_incidents_2020}.}
    \end{subfigure}
    \\
    \begin{subfigure}[a]{0.9\textwidth}
        \includegraphics[width=0.45\textwidth]{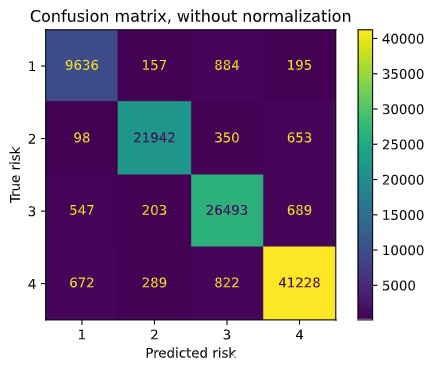}
        ~
        \includegraphics[width=0.45\textwidth]{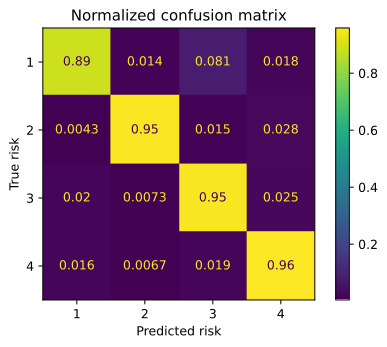}
        \caption{Results obtained using our data preparation and feature selection methodology.}
    \end{subfigure}
    \caption{Confusion Matrix for the threat level prediction use case.}
    \label{fig:confusionmatrix}
\end{figure}

Training a model with almost 1M attributes from MISP containing 39 features, we can predict the threat level of their related events with a prediction accuracy of 96\%, as reported in Figure~\ref{fig:confusionmatrix}(b). Following the same feature selection and data preparation as in~\cite{fotiadou_incidents_2020}, we obtain a lower accuracy (91.8\%) (Figure\ref{fig:confusionmatrix}(a)).\footnote{The authors of~\cite{fotiadou_incidents_2020} compare multiple methods with different accuracy levels, where CNNs provide an accuracy of up to 98\% with their dataset. Nevertheless, the results are not directly comparable, as the used datasets differ in the available features and in the nature and distribution of the events.}

We are confident that this score can be greatly improved with better feature engineering and model optimization, but this is not the focus of this work. As a matter of fact, these results show that the impact of the secure workflow in the accuracy of the computation is negligible for the models we are considering, as we obtain the same accuracy results with and without encryption.

In order to understand the importance of each feature in the model, we calculated their relative weight in the trained regression model. This also helps us evaluate the importance of data that may have different confidentiality levels and determine whether they are actually needed for maximizing the prediction accuracy. Most of these attributes are Boolean values indicating the presence or absence of certain metadata. We take the resulting weights from the trained regression and analyze the impact and significance of each of the attributes used in training in the classifier accuracy. The features that present a higher correlation with the assigned label have a larger weight in the regression, and vice-versa. The most relevant features together with their associated weights are reported in Table~\ref{tab:topmispfeatures}.

It can be seen that the most relevant features, such as the date or whether the event was published in MISP, do not represent confidential information and are specific to the MISP entry rather than to the incident itself.

\begin{table}
    \caption{Top 5 fields identified by the secure regression analysis and their associated weights.}
    \label{tab:topmispfeatures}
\centering
    \subcaption{Results obtained using our data preparation and feature selection methodology.}
    \begin{tabularx}{.9\textwidth}{|p{.12\textwidth}|l|p{.64\textwidth}|}\hline
        \textbf{Name}&	\textbf{Weight}&\textbf{Description}\\\hline
        MD5&	5.25&	Presence of an MD5 hash related to the event. This usually indicates that the event relates to an identified malware, hence the threat level is known.\\\hline
Analysis&	4.84&	Indicates whether the analysis is completed.\\\hline
IoC information&	3.76&	Presence of information about the Indicator of Compromise.\\\hline
Distribution&	3.10&	Whether the event has been distributed to other participating organizations of the network.\\\hline
Report information&	1.96&	Presence of report information about the event.\\\hline
    \end{tabularx}
    \subcaption{Results obtained using the methodology in~\cite{fotiadou_incidents_2020}.}
    \begin{tabularx}{.9\textwidth}{|p{.12\textwidth}|l|p{.64\textwidth}|}\hline
        \textbf{Name}&	\textbf{Weight}&\textbf{Description}\\\hline
        Week&	5.09&	Week of the event.\\\hline
Month&	4.99&	Month of the event.\\\hline
Published&	2.03&	Whether the event was published in MISP.\\\hline
Distribution&	1.07&	Whether the event has been distributed to other participating organizations of the network.\\\hline
Attribute count&	1.01&	Number of attributes of the event.\\\hline
    \end{tabularx}
\end{table}

\subsubsection*{Discussion}
These results prove the feasibility of the use of machine learning algorithms to reliably predict the threat level of MISP events, and also the negligible accuracy impact of doing so in a privacy preserving manner. The used training process can be transferred to a case that has confidential and/or critical non-shareable data as inputs, and the results can be generalized to automatic prediction using confidential features.

The nature of the used fields is very specific to the input events, rather than to the reported incident. Therefore, the predicted risk is evaluated from the context of previously entered events. This could potentially help an operator to automate tasks but should not be seen as a risk evaluation of the incident. Our MISP instance uses the public feeds provided by CIRCL, and compared to the work by Fotiadou et al.~\cite{fotiadou_incidents_2020} we do not have the same fields, as some are related to attributes that are specific to a certain dataset. This implies a challenge to reproduce the exact same results in~\cite{fotiadou_incidents_2020}.%, as not all the fields used in~\cite{fotiadou_incidents_2020} are available in our MISP dataset: Some are related to attributes that are specific to a certain dataset and not present in the CIRCL MISP.

As already mentioned, we used information that is already shared in the CIRCL MISP, but we argue that the fields that are identified as more representative of the threat level of an event in a public MISP database are not automatically generated by the system, but entered by the operator analysis. This makes the automation of such a system impractical. This observation is closely related to the nature of the information shared in the used public MISP dataset: Most fields are manual reports of indicators of compromise.

To overcome this, events entered in the MISP instances used for the secure computation should contain more standardized information about the system that detects the intrusion, which can be achieved by directly populating MISP from rules extracted from a SOC or SIEM. Our system removes the necessity of sharing the events in clear with the other participants and therefore enables such systems to automatically operate without risking unsupervised leaks of confidential information.

\subsection{Training and prediction of DDoS attacks}
In the study~\cite{batchu_generalized_2021}, the authors develop and compare several machine learning methods: Logistic Regression (LR), Gradient Boosting (GB), Decision Tree (DT), k-nearest neighbors (KNN), and Support Vector Machines (SVM). Their goal is to optimize the classification accuracy and optimize the detection of malign traffic. As mentioned above, for our analysis we focus on a logistic regression as a common exemplifying and interpretable baseline to enable fair comparisons between the different approaches.

The CICDDoS2019 dataset used in~\cite{batchu_generalized_2021} is a traffic capture done over two different days. For each day, the dataset comprises captures of traffic from multiple services, namely LDAP, MSSQL, NetBIOS, Portmap, Syn, UDP, and UDPLag. All these captures have 88 common recorded fields. Therefore, we can aggregate in one dataset the captures from the different protocols by taking these 88 fields. Aggregated and converted into an uncompressed CSV, the datasets have the sizes shown in Table~\ref{tab:CICDDDoS2019sizes}.

\begin{table}
    \caption{Size of the CICDDoS2019 datasets used in~\cite{batchu_generalized_2021}.}
    \label{tab:CICDDDoS2019sizes}
    \centering
    \begin{tabular}{|l|l|l|}\hline
        \textbf{Dataset Name}&	\textbf{Dataset Size}&\textbf{Number of rows}\\\hline
        DAY1&	8.2 GB&	20.36 M\\\hline
        DAY2&	21 GB&	50.03 M\\\hline
    \end{tabular}
\end{table}

The two days, namely day 1 (03-11) and day 2 (01-12) are independent captures that can be used as training and test sets. Batchu and Seetha~\cite{batchu_generalized_2021} separated their analysis in four cases:

\begin{itemize}
\item Case 1: Splitting the day 1 dataset: 70\% for training, 30\% for testing.
\item Case 2: Using day 1 for training and day 2 for testing.
\item Case 3: Splitting the day 2 dataset: 70\% for training, 30\% for testing.
\item Case 4: Using day 2 for training and day 1 for testing.
\end{itemize}

For the following analysis, we focus on Case 1, using the Day 1 dataset. Data preparation steps have been reproduced following the methods used in the article: Conversion to numerical data, removal of duplicates, replacement of null values by the mean, and standardization of the dataset. But the impact of data imbalance and the techniques used in~\cite{batchu_generalized_2021} to address it deserve a separate analysis, that we detail in the following section.

\subsubsection{The impact of data imbalance and synthetic data generation strategies}
An important aspect of this dataset to take into account for training is that it is strongly imbalanced, with a ratio of 0.000206 malignant vs benign network requests. This is typical for such real-world cases, because there is usually much more benign than malign traffic.
In order to properly train a logistic regression, the data needs to be rebalanced. A natural approach would be to randomly remove elements from the over-represented class from the imbalanced dataset in order to reach parity, but this approach vastly reduces the dataset size. The approach taken by Batchu and Seetha~\cite{batchu_generalized_2021} is to rebalance the data using synthetic data generation of the minority class, based on the initial dataset. We have also used this oversampling method, denoted SMOTE (Synthetic Minority Oversampling Technique), that rebalances the dataset to a 50\% split for the two classes by adding synthetic data close to the neighboring values of the minority class. This technique greatly improves the performance of the trained model, but artificially shapes the data in a very homogeneous and predictable way, which has serious consequences on the generalizability and robustness of the trained model.

In general, using synthetic data with such an imbalanced dataset in order to mitigate the lack of sufficient malign samples can be problematic. Indeed, a database with a large amount of synthetic data will make the model adapt to the synthetic data and the rules/patterns used to generate it, which can bias the model to converge towards the synthetic generation rule instead of the actual distinguishing features of the malign data. Therefore, we conjecture that the use of the SMOTE oversampling creates an artificially homogeneous dataset that introduces an overfitting of the model to the synthetically generated data and will result in an artificially high accuracy.

We verified this assumption with an empirical experiment. First, we define two datasets:
\begin{itemize}
\item \emph{Real}: a dataset that contains all the rare events and that is balanced with a random subsampling of the common events.
\item \emph{Synthetic}: a dataset that contains the original unbalanced data, after (i) balancing it by means of oversampling the rare events with SMOTE and then (ii) subsampling by randomly removing events until the set reaches the same size of the \emph{Real} dataset. The subsampling is done class-wise, ensuring that the subsampled dataset is still balanced.
\end{itemize}
We then repeated the following experiment 10 times and averaged the obtained results:
\begin{enumerate}
\item Generate new \emph{Real} and \emph{Synthetic} datasets.
\item Randomly split the generated datasets into 70/30 training/test sets.
\item Train a classifier on each dataset.
\item Evaluate the classifiers.
\end{enumerate}

Table~\ref{tab:realsynthaccuracy} reports the accuracy matrix of the experiment, as well as the standard deviation of the different runs.

\begin{table}
    \caption{Accuracy obtained when training and evaluating on sets taking from the two datasets \emph{Real} and \emph{Synthetic}. Accuracy is represented with average $\pm$ standard deviation.}
    \label{tab:realsynthaccuracy}
    \centering
    \begin{tabular}{|l|l|l|}\hline
        %\textbf{Train | Test}
        &	Test \emph{Real}&Test \emph{Synthetic}\\\hline
        Train \emph{Real}&	0.99501 ± 2.357e-5&	0.99858 ± 7.552e-5\\\hline
        Train \emph{Synthetic}&	0.9904 ± 3.640e-4&	0.9984 ± 1.324e-4\\\hline
    \end{tabular}
\end{table}

We observe that the synthetic test dataset always gives better accuracy, regardless of the dataset used for training. This supports the hypothesis that the rules used to generate the synthetic data are simpler than the actual data, hence it does not properly convey the relevant features of the real data. Moreover, the accuracy is always higher than when evaluated on the real dataset. When evaluated on the real test data, the classifier trained on real data performs better than the classifier trained on synthetic data. Moreover, the classifier trained on synthetic data has a much worse performance when evaluated on real data. This is another piece of evidence supporting the hypothesis that the synthetic data is ``simpler'' than the actual real data. Finally, the standard deviation is an order of magnitude smaller than the differences between different accuracy levels in each dataset, so the results are statistically significant.

This issue is unfortunately very common in data science; the need to perform oversampling comes from a lack of data or evidence of the malign case. The best way to improve this limitation is to be able to collect more data, from the same source or from other sources. This is where our solution becomes a key asset, by enabling the use of a larger data pool instead of resorting to oversampling, therefore improving the quality of the trained models and their robustness against real attacks. For instance, an organization that has not yet been victim of a DDoS attack can make use of the data from other organizations that have this experience and therefore preemptively improve its defense system.

\subsubsection{Accuracy evaluation}
In order to reproduce the results from Batchu and Seetha~\cite{batchu_generalized_2021}, we first re-implemented the method in Go with the centralized data and processed it in cleartext to obtain the accuracy baseline; we also used a cleartext decentralized implementation to determine the effect of distribution.

As noted above, the dataset was initially highly imbalanced and comprised a relatively low amount of malign traffic instances. Batchu and Seetha~\cite{batchu_generalized_2021} used the SMOTE method to rebalance the dataset, greatly improving the apparent model performance.

We now replicate the obtained results when data cannot be centralized, with multiple data providers training a collective model by making use of their combined datasets without centralizing the data. For this purpose, we split the original dataset into equally sized subsets, one for each node, and compare the performance of the model when (a) trained with a centralized approach versus (b) when the nodes perform the training only with their local samples or (c) with a distributed training.

\begin{figure}
\centering
    \begin{subfigure}[a]{0.41\textwidth}
    \includegraphics[width=\textwidth]{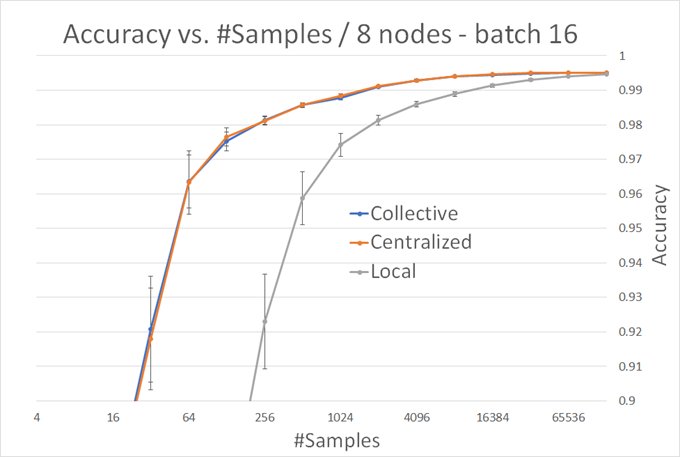}
    \caption{Accuracy for different sizes of the global dataset. Error bars indicate a 95\% confidence interval}
    \end{subfigure}$\;$
    \begin{subfigure}[a]{0.54\textwidth}
    \includegraphics[width=\textwidth]{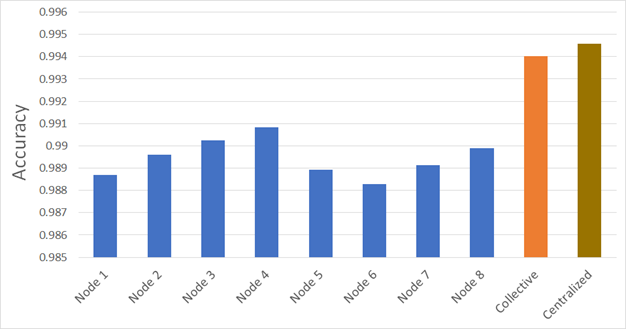}
    \caption{Local accuracy at each node for one randomly chosen data distribution when compared with the achieved collective accuracy.}
    \end{subfigure}
    \caption{Accuracy of the collective (8 nodes), central and local training.}
    \label{fig:DDoSaccuracy}
\end{figure}

Figure~\ref{fig:DDoSaccuracy}(a) shows the accuracy of the centralized, distributed (collective) and local trained models, averaged over 20 runs, for an exponentially increasing number of total samples. We observe that when the data is scarce, the centralized and decentralized training provide a significantly higher accuracy than the local training; the variance of the average accuracy in those cases becomes negligible much faster than in the local training. Moreover, decentralized training is always very close to centralized training. Eventually, when the dataset is large enough, all methods converge to the same accuracy, which is expected.

Figure~\ref{fig:DDoSaccuracy}(b) shows the results for a single run, where each node has 1,000 samples, as a particular example of the performance at each node compared with the achievable (ideal) results when data is centralized, and how this centralized accuracy is almost perfectly recovered when running a distributed protocol without centralizing the data.

\subsubsection{Performance evaluation}
In order to evaluate the overall performance and computation complexity of the proposed system, we performed additional experiments with different input sizes, where we used a standard off-the-shelf server virtual machine (VM) with 16 virtual cores at 2.1GHz and 64 GB of RAM.

\begin{figure}
\centering
    % \includegraphics[width=\textwidth]{fig_8_training_time.png} 
    % from https://docs.google.com/spreadsheets/d/1UWZCpAsE5FYNf5glNty0WwitGkEhl69I75diIIsdjxw/edit#gid=46831662
    
    \includegraphics[width=\textwidth]{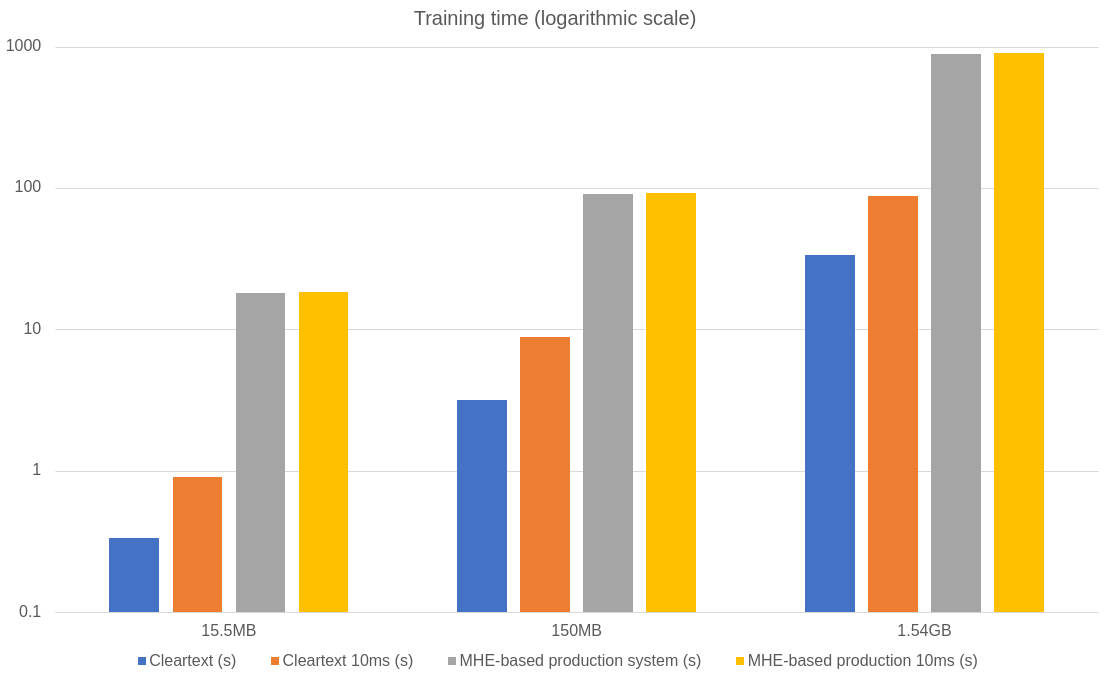} 
    % Updated from https://1drv.ms/x/s!AoPZUIgXoAvkc6jh9Y26azo1MYU?e=GdWR2g
    \caption{Cleartext vs. encrypted distributed training time (only computation) for growing dataset sizes, including executions with a simulated 10~ms network delay.}
    \label{fig:DDoSperformance}
\end{figure}

During the experiment, the peak RAM usage for the encrypted solution was 13 GB, mostly due to the initial generation of keys for the 10 nodes. In an operational setting, those keys are generated once among the nodes during the network setup phase, and stored on disk most of the time. The training time was recorded with the data split in 10 virtual nodes both for the encrypted and cleartext version with no network delay. We observe that the encrypted training time increases linearly with the dataset size, and it represents a constant overhead versus the cleartext training computing time, as shown in Figure~\ref{fig:DDoSperformance}.

This figure also shows the results when simulating a network delay of 10~ms, which is representative of a LAN network. In fact, communication time (transmission and latency) is usually the bottleneck in this kind of distributed settings, and that normally neglects the computation overhead between the cleartext and encrypted approaches. For this case, we can quantify the used bandwidth by computing the communicated information during the exchange of ciphertexts at each iteration of the learning algorithm, as this is the dominant factor.
Globally, this communication can be quantified as: $\#Local\_Iterations \times \#Nodes \times Ciphertext\_Size \times 1.5$ bytes, where the ciphertext size equals $N \times \#q_i \times 2 \times 8$ bytes. $N$ and $q_i$ are cryptographic parameters that, 
for 128-bit post-quantum security, will be chosen as $N\in\{16387, 32768, 65536\}$, and $\#q_i \in [1,30]$. This implies a communication ranging between 0.26 MB to 31.4 MB per node and iteration, depending on the choice of the parameter set and the computation to run.

From this calculation and from the results shown in Figure~\ref{fig:DDoSperformance}, it can be seen that the used implementation (MHE-based production system) is always within 1-order of magnitude of the cleartext solution when the network latency is 10~ms, and this gap gets further reduced when the network latency grows (for WAN settings). %presents numerous improvements and optimizations over the original academic implementation of the regression training system [4], which lead to a reduction in execution time that ranges between one and two orders of magnitude, as shown in the gray bars in Figure~\ref{fig:DDoSperformance}.

Also, it must be noted that normally processes can be parallelized both within the cores of a server and horizontally scaled across multiple nodes. The proposed encrypted solutions are highly parallelizable, so they can naturally take advantage of the available cores and nodes to further reduce the overall response time.

\subsubsection{Discussion}
Oversampling is typically used when the available data is insufficient; this is the case with unbalanced datasets where the to-be-detected events are present in an extremely low ratio. Nevertheless, oversampling can introduce bias and overfitting of the trained model, artificially raising the obtained accuracy and producing unrealistic and unusable results that cannot be reproduced on a larger real dataset, as the resulting model has poor generalizability. In these cases, the only alternatives are to source further data or, when centralization is not possible due to confidentiality issues, to use a distributed secure approach like the one we propose. This is indeed the case with the CICDDoS2019 dataset and, in general, with network captures of DDoS attacks: The dataset is highly imbalanced, and each organization has very little evidence of malign traffic. Hence, they would benefit even more from using the collective evidence of those events and computing a global model with as many participants as possible.

Our tests on the CICDDoS2019 dataset validate the aforementioned conjecture. This reinforces the applicability and usefulness of our proposed distributed method. Furthermore, the comparative evaluation of centralized and distributed training also shows the benefits and raw accuracy gain that cyberdefense teams can achieve by engaging into collaborative analyses following our proposed mechanisms, without the need to transfer or disclose their (potentially confidential) data.

\section{Discussion and Extensions}
Our framework offers a novel solution to reduce the tension organizations face when managing data protection risks while benefiting from information sharing. All in all, our contribution extends the current scientific literature on CTI by mitigating the privacy / utility trade-off in cybersecurity information sharing.

In this work, our framework was deployed and tested in an environment that uses MISP data. While this might be considered as a limitation, we believe that our approach is generalizable to other "sharing organizations", such as Security Operations Centers (SOCs), Computer Emergency Response Team (CSIRTS), or Information Sharing and Analysis Centers (ISACs). Currently, our solution is also being tested in the health sector and in the banking and financial sector.% in the context of a closed Financial Services Information Sharing and Analysis Center (FS-ISAC).

\subsection{Policy Implications for CTI Sharing}

Our framework may also enhance legislative initiatives that support (or sometimes make mandatory) cybersecurity information sharing, in order to produce centralized CTI. For instance, in 2015, the US ``Cybersecurity Information Sharing Act''\footnote{\url{https://www.congress.gov/bill/114th-congress/senate-bill/754}} encouraged the sharing of CTI indicators between government and private organizations. This act required the US federal government to facilitate and promote CTI sharing, including the sharing of ``classified and declassified cyber threat indicators in possession of the federal government with private entities [...]''. In 2016, the National Institute of Standards and Technology (NIST) issued a publication (NIST SP 800-150)\footnote{\url{https://nvlpubs.nist.gov/nistpubs/SpecialPublications/NIST.SP.800-150.pdf}} which further outlined the necessity for CTI Sharing as well as a framework for implementation, which is interoperable with the framework presented in this article.

In the EU, the article 33 of the GDPR\footnote{\url{https://gdpr-info.eu/art-33-gdpr}} requires organizations that experience a personal data breach to report it within 72 hours to authorities. Our framework could support this process, without necessarily having to share all the information containing sensitive data on customers or suppliers, which is often the case in the banking or defense sectors. Additionally, when dealing with data collaborations on cyber intelligence that can be considered personal data, the proposed system can also constitute the appropriate measures conducive to the compliance of GDPR articles 25 and 32 thanks to its privacy- and security-by-design properties, as well as providing the necessary supplementary measures for enabling cross-border collaborations between EU and non-member states (international data transfers), according to GDPR article 46~\cite{corrales21}.

Further research will be very much needed to shed light onto the effectiveness of such laws, as some current policy recommendations support the view that the effectiveness of centralized mandatory security-breach reporting to authorities is limited ~\cite{mermoud_three_2019}. Our approach may offer an interesting alternative to such centralized and government-driven information sharing.

%To conclude, our framework provides a concrete novel solution for orchestrating collaborative cybersecurity. In our modern world, no single organization can provide for its security or data protection in isolation from all others. Our solution mitigates the free-rider problem, by getting around human bias and the problem of misaligned incentives. In this sense, our work fills an important gap in the existing literature ~\cite{wagner2019cyber} on cybersecurity information sharing.

\subsection{Identity management and authentication and authorization infrastructure}
\label{sec:keycloak}
The authentication and authorization in our platform rely on an external component that is integrated with our system: Keycloak. It is an identity and access management system that supports a wide array of authentication and authorization protocols (e.g., OpenID Connect, OAuth2, SAML) and users directories (e.g., LDAP, Active Directory). The use of Keycloak in our infrastructure enables a broad compatibility with existing systems.

The authentication standard used by our architecture is OpenID Connect. Each node of the network establishes a trust relation with one or more instances of Keycloak using JWKS (JSON Web Key Set) to retrieve the public signing key (JWK) of the instance(s). Using these keys, each node is able to verify the authenticity of any message emitted by the Keycloak instance(s).

The users’ identities are managed by Keycloak, thus to interact with a node the user first needs to authenticate against Keycloak, and obtain a signed JWT (JSON Web Token) that contains the identity and authorizations of the user. This JWT is embedded in all requests made to the node by the client (in the HTTP header as a bearer authorization), and used to enforce the authentication and authorizations managed by the Keycloak instance that emitted the JWT.

From this basic infrastructure, the setup is flexible and is only a matter of deployment and configuration, with the following options:
\begin{itemize}
\item A unique centralized Keycloak instance can be used across the network by all nodes. While this is easier to operate, it introduces a single point of failure.
\item Alternatively, a federated approach can be used: Each node can deploy its own Keycloak instance to manage its users. One of the following two operating modes can then be used:
    \begin{itemize}
    \item Each node authenticates its own clients, and it trusts the other nodes to authenticate their own clients properly, thus trusting incoming computation requests from other nodes. 
    \item Each node authenticates its own clients and also authenticates the clients of other nodes by using the Keycloak instances of those nodes.
    \end{itemize}
\end{itemize}

Note that a hybrid approach with some nodes deploying their own instance and some other nodes relying on the instance of another node is also doable.

Configuration-wise, each Keycloak instance can be configured with any kind of user source for authentication. It can be its own manually-operated user-source, or use an external one that already exists. This can be a simple user directory like LDAP or Active Directory, or a full-blown protocol like OpenID Connect, OAuth2 or SAML in case there is an existing auth infrastructure.

Then, a policy for authorizations and role-management needs to be defined and configured in Keycloak. Our architecture defines a number of different roles according to the used data and the performed computations, that need to be set for each use. This can be done manually or automatically; e.g., with role-mapping rules based on belonging to certain groups of the organization.

\subsection{Malicious adversaries}
Our assumed threat model considers passive adversaries with up to N-1 colluding parties, as this is a reasonable assumption for competitive but collaborative institutions, that contribute to the network in a win-win scenario, and for which a malicious behavior would severely harm their reputation and provoke their banishment from the network.
When faced with scenarios where it is reasonable to assume active (malicious) adversaries, our system could be extended to an active-adversarial setting by using standard verifiable computation techniques, e.g., resorting to zero-knowledge proofs and redundant computation. This would, though, come at the cost of an increase in the computational complexity.
Finally, it is worth noting that the main confidentiality target sought by this work relates to protection of the data during computation, against the other nodes in the network and/or the computing infrastructure. This is achieved by minimizing the released data only to the computation results that are required by the system functionality, and by limiting this disclosure to only the authorized querier. It is possible to add an additional layer of protection to the released results when the querier can also be considered curious or malicious, by resorting to obfuscation techniques tied to privacy frameworks such as differential privacy~\cite{shokri15}. These approaches distort the outputs of the computation according to a determined privacy budget associated to the querier, hence introducing a privacy-utility trade-off (increased privacy implies reduced data utility). These techniques are orthogonal and complementary to the ones used in this work. Moreover, when the proposed secure computation techniques are used in combination with differential privacy, the leakage minimization of the former enables an optimal privacy-utility trade-off on the latter.

\section{Conclusions}
The need for leveraging the combined Cyber threat intelligence (CTI) data of organizations is becoming increasingly relevant in a scenario of fast digital transformation that poses new challenges and threats.
The communities around the Malware Information Sharing Platform (MISP) are growing but are limited to sharing only the least sensitive datasets they collect. The increased amount of data that has to be secured is leading to threat intelligence sharing bottlenecks.
Leveraging the already existing sharing communities around MISP, we propose a solution to enable participant organizations to make use of their sensitive cyber threat intelligence by providing a framework and scalable software capable of orchestrating successful collaborations leading to statistically significant and useful insights that can support and eventually improve the effectiveness and reliability of the implemented cyberdefense processes.

We have identified three representative use cases that would inform the participating organizations about both key metrics on the gathered threat intelligence and provide powerful analytical tools to be alerted and take action on incoming threats. Among these, we have analyzed a DDoS detection case and showed the limitations of the oversampling used when the available data about malign events is insufficient, evidencing that distributed training is the only viable solution when centralization is not possible due to the aforementioned confidentiality constraints.

The proposed solution has been implemented and has reached sufficient maturity to be brought to an exploitable product that goes beyond the academic phase, and it is being tested in real pilots involving critical infrastructures such as hospitals.

To conclude, our framework provides a concrete novel solution for orchestrating collaborative cybersecurity. In our modern world, no single organization can provide for its security or data protection in isolation from all others. Our solution mitigates the free-rider problem, by getting around human bias and the problem of misaligned incentives. In this sense, our work fills an important gap in the existing literature ~\cite{wagner2019cyber} on cybersecurity information sharing.
%\begin{table}
%  \caption{Frequency of Special Characters}
%  \label{tab:freq}
%  \begin{tabular}{ccl}
%    \toprule
%    Non-English or Math&Frequency&Comments\\
%    \midrule
%    \O & 1 in 1,000& For Swedish names\\
%    $\pi$ & 1 in 5& Common in math\\
%    \$ & 4 in 5 & Used in business\\
%    $\Psi^2_1$ & 1 in 40,000& Unexplained usage\\
%  \bottomrule
%\end{tabular}
%\end{table}

%\begin{figure}[h]
%  \centering
%  \includegraphics[width=\linewidth]{sample-franklin}
%  \caption{1907 Franklin Model D roadster. Photograph by Harris \&
%    Ewing, Inc. [Public domain], via Wikimedia
%    Commons. (\url{https://goo.gl/VLCRBB}).}
%  \Description{A woman and a girl in white dresses sit in an open car.}
%\end{figure}

%\begin{acks}
%  Identification of funding sources and other support, and thanks to individuals and groups that assisted in the research and the preparation of the work should be included in an acknowledgment section, which is placed just before the reference section in your document.
%\end{acks}

\bibliographystyle{plain}%ACM-Reference-Format}
\bibliography{biblio}

\appendix

\section{Summary of features present on the CIRCL MISP instance}
In this appendix, we detail all the features present in our replica of the CIRCL MISP, listed in Table~\ref{tab:allcirclemispfeatures}.

\begin{table}[h!]
\centering
\caption{List of features available in the CIRCL MISP instance used in the threat level prediction use case.}
\label{tab:allcirclemispfeatures}
\begin{tabular}{|c|c|c|}
\hline
% y &
% h	&
% week	\\ \hline
% e\_y &
% e\_h	&
% e\_week	\\ \hline
% https &
% version	&
% statically	\\ \hline
% MSB &
% executable	&
% Malicious	\\ \hline
% China &
% Intel	&
% virustotal	\\ \hline
% dotcom &
% dotorg	&
% dotpl	\\ \hline
% has\_comment &
% info\_malware	&
% info\_IOC	\\ \hline
% info\_report &
% info\_delta	&
% info\_MD5	\\ \hline
% info\_endpoint &
% info\_Trickbot	&
% info\_Cryptolaemus	\\ \hline
% info\_Campaign &
% info\_Emotet	&
% info\_Cryptolaemus	\\ \hline
% info\_Incremental &
% info\_Ransomware	&
% info\_Malspam	\\ \hline
% info\_attacks &
% info\_Phishing	&
% info\_activity	\\ \hline
% tag &
% object\_relation	&
% category	\\ \hline
% type &
% comment	&
% published	\\ \hline
% analysis &
% attribute\_count	&
% distribution	\\ \hline
% sharing\_group\_id &
% publish\_timestamp	&
% sighting\_timestamp	\\ \hline
% threat\_level\_id & & \\ \hline
attribute\_year &
attribute\_hour &
attribute\_week \\ \hline
event\_y &
event\_h &
event\_week \\ \hline
https &
MSB &
executable \\ \hline
Malicious &
China &
Intel \\ \hline
virustotal &
dotcom &
dotorg \\ \hline
dotpl &
has\_comment &
info\_malware \\ \hline
info\_IOC &
info\_report &
info\_delta \\ \hline
info\_MD5 &
info\_endpoint &
info\_Trickbot \\ \hline
info\_Cryptolaemus &
info\_Campaign &
info\_Emotet \\ \hline
info\_Incremental &
info\_Ransomware &
info\_Malspam \\ \hline
info\_attacks &
info\_Phishing &
info\_activity \\ \hline
category &
published &
analysis \\ \hline
attribute\_count &
distribution &
sharing\_group\_id \\ \hline
\end{tabular}
\end{table}

%%%% unformated list
% y
% h
% week
% e_y
% e_h
% e_week
% https
% version
% statically
% MSB
% executable
% Malicious
% China
% Intel
% virustotal
% dotcom
% dotorg
% dotpl
% has_comment
% info_malware
% info_IOC
% info_report
% info_delta
% info_MD5
% info_endpoint
% info_Trickbot
% info_Cryptolaemus
% info_Campaign
% info_Emotet
% info_Cryptolaemus
% info_Incremental
% info_Ransomware
% info_Malspam
% info_attacks
% info_Phishing
% info_activity
% tag
% object_relation
% category
% type
% comment
% published
% analysis
% attribute_count
% distribution
% sharing_group_id
% publish_timestamp
% sighting_timestamp
% threat_level_id

\end{document}